# Modelling the impact of deep-water crustacean trawl fishery in the marine ecosystem off Portuguese Southwestern and South Coasts: I) the trophic web and trophic flows

Maria Angeles Torres[a,*], Paulo Fonseca[a,b], Karim Erzini[a], Teresa Cerveira Borges[a], Aida Campos[a,b], Margarida Castro[a], Jorge Santos[c], Maria Esmeralda Costa[a], Ana Marçalo[a], Nuno Oliveira[d], José Vingada[e]

[a] Centro de Ciências do Mar, Universidade do Algarve, Campus de Gambelas, 8005-139 Faro, Portugal

[b] Instituto Português do Mar e da Atmosfera. Av. Brasília, 1449-006 Lisboa, Portugal

[c] Norwegian College of Fishery Science, University of Tromsø, NO-9037, Tromsø, Norway

[d] Portuguese Society for the Study of Birds, Avenida João Crisóstomo, 1000-179 Lisboa, Portugal

[e] Sociedade Portuguesa de Vida Selvagem, Portugal

* Corresponding author: Maria Angeles Torres matorres@ualg.pt, torresleal.ma@gmail.com

**Abstract**

The concentration of the population in coastal regions, in addition to the direct human use, is leading to an accelerated process of change and deterioration of the marine ecosystems. Human activities such as fishing together with environmental drivers (e.g. climate change) are triggering major threats to marine biodiversity, and impact directly the services they provide. In the South and Southwest coasts of Portugal, the deep-water crustacean trawl fishery is not exemption. This fishery is recognized to have large effects on a number of species while generating high rates of unwanted catches. However, taking into account an ecosystem-based perspective, the fishing impacts along the food web accounting for biological interactions between and among species caught remains poorly understood. These impacts are particularly troubling and are a cause of concern given the cascading effects that might arise. Facing the main policies and legislative instruments for the restoration and conservation of the marine environment, times are calling for implementing ecosystem-based approaches to fisheries management. To this end, we use a food web modelling (Ecopath with Ecosim) approach to assess the fishing impacts of this particular fishery on the marine ecosystem of southern and southwestern Portugal. In particular, we describe the food web structure and functioning, identify the main keystone species and/or groups, quantify the major trophic and energy flows, and ultimately assess the impact of fishing on the target species but also on the ecosystem by means of ecological and ecosystem-based indicators. Finally, we examine limitations and weaknesses of the model for potential improvements and future research directions.

## 1. Introduction

Bottom trawling is recognized to be one of the most damaging and less environment friendly fisheries in the world (Jackson *et al.*, 2001). This practise poses major risks on the marine ecosystems including for example alteration of habitat structure, damage to bottom communities, loss of biodiversity, structure and productivity of benthic communities (Gray, 1997; Jennings and Kaiser, 1998; Pauly *et al.*, 2002). These changes are ultimately translated into population declines (Lewison *et al.*, 2004) and food web alteration that can affect ecosystem functions (Pauly *et al.*, 1998). The removal of large, slow growth and long-lived predatory fish because of current fishing practices is affecting the reproductive characteristics of stocks even the structure of the whole ecosystem (Myers and Worm, 2003; Worm *et al.*, 2005). In addition, fishing activities generate high rates of unwanted species which represents a wasteful use of living resources (Alverson *et al.*, 1994; Kelleher 2005; Zeller and Pauly, 2005; Ulhmann *et al.*, 2013) with important ecological consequences (Tsagarakis *et al.*, 2014; Heath *et al.*, 2014) and changes at the ecosystem level which may in turn affect fisheries sustainability (Zhou *et al.*, 2010).

In Portuguese waters (ICES subdivision IXa), the deep-sea crustacean trawl fishery is not an exception. This multi-species fishery constitutes a very important part of the fishing fleet particularly in the Algarve region (Costa *et al.*, 2008). The main fishing grounds are located at the shelf edge and continental slope off the southern and southwestern coasts of Portugal at depths of 200-700 m. The main target species rose shrimp (*Parapenaeus longirostris*), Norway lobster (*Nephrops norvegicus*) and red shrimp (*Aristeus antennatus*) inhabit soft habitats composed of a mixture of mud and silt (Monteiro *et al.*, 2001; Campos *et al.*, 2007). Other deep-water crustacean species, namely the purple shrimp (*Aristaeomorpha foliacea*) and the scarlet shrimp (*Aristaeopsis edwardsiana*), are only occasionally targeted or incidentally caught at depths greater than 500 m (Silva *et al.*, 2015). This multi-species fishery operates by using codend mesh sizes of 55 mm and 70 mm and targets primarily on the commercial species according to availability, market demand and price (Silva *et al.*, 2015).

Although principally targeting commercially-valuable crustacean species, this multi-species fishery has a negative environmental impact as it generates substantial by-catch and discard rates (34-70%) (Monteiro *et al.*, 2001; Erzini *et al.*, 2002; Castro *et al.*, 2005; Costa *et al.*, 2008). The most represented commercial by-catch fish species include blue whiting (*Micromesistius poutassou*), undersized hake (*Merluccius merluccius*), blue jack mackerel (*Trachurus picturatus*) and Atlantic horse mackerel (*Trachurus trachurus*) (Costa *et al.*, 2008). Benthic sharks are also accidentally caught by this fishery (Moura *et al.*, 2018). However, discards of target species rose shrimp and Norway lobster are negligible (Fernández *et al.*, 2015).

Historical landings, economic trends and opportunities for the Portuguese fisheries sector have been fully documented in the area (e.g. Hill and Coelho, 2001; Leitão *et al.*, 2014; Leitão, 2015; Leitão and Baptista, 2017). Trends in landings related to environmental variables have been analysed (Erzini, 2005; Teixeira *et al.*, 2014; Gamito *et al.*, 2015, 2016) and changes in the trophic level of such landings have been reported (Baeta *et al.*, 2009). Although these studies evidence potential effects in the marine ecosystem structure and functioning, the impact of deep-water crustacean trawl fishery along the food web accounting for biological interactions between and among species caught remains poorly understood. These impacts are particularly troubling and are a cause of concern given the cascading effects that might arise.

The year 2020 is a landmark for the main policies and legislative instruments for the restoration and conservation of the marine environment in Europe (e.g. Common Fisheries Policy (CFP), Marine Strategy Framework Directive (MSFD), EU Biodiversity Strategy, Water Framework Directive (WFD), Marine Protected Areas (MPA), Convention on Biological Diversity-Aichi targets). Hence, it is timely to make operational an ecosystem-based approach to fisheries management (EBFM) (Garcia *et al.*, 2003) that is the most suitable tool to integrate sustainable exploitation activities consistent with the existence of ecological interactions, environmental constraints, protection of natural and cultural heritage, and restoration of marine ecosystems (Link and Browman, 2014). Although the theory behind EBFM is well developed, its full implementation still lags behind (Berkes, 2012).

Within the context of EBFM, the Ecopath with Ecosim (EwE) approach is by far the most commonly used ecosystem modelling platform to create ecological models (Polovina, 1984; Christensen and Pauly, 1992; Christensen and Walters, 2004, 2011). This approach integrates human activities (i.e. fisheries) within an ecosystem context and evaluating their impacts on marine food webs, including trophic interactions between and among species and environmental factors (Coll and Libralato, 2012). Outcomes of this model provide insights into ecosystem structure and functioning, adding to our understanding the control of energy flows transferred through the food web while allowing scientists and managers to modify components of the ecosystem to explore past and future impacts of fishing and environmental disturbances as well as to explore optimal fishing policies. This package also gathers a significant amount of information at an ecosystem level based on network analysis and information theory (Ulanowicz, 1986; Christensen and Pauly, 1993; Heymans and Baird, 2000).

The primary goal of our study is to assess for the first time the impact of deep-water crustacean trawl fishery on the upper trophic levels and explore the effects cascading down the food web to small fish, using a food web modelling approach (Ecopath with Ecosim). In particular, we characterize the structure and resilience of the southern Portuguese marine ecosystem by describing the food web structure and functioning, identifying the main keystone species and/or groups of the ecosystem, quantifying the main trophic flows and energy transferences and, assessing the impact of fishing on the target species but also on the ecosystem by means of ecological and ecosystem-based indicators. Finally, we examine limitations and weaknesses of the model for potential improvements and future research directions.

## 2. Study site

The study area, where the deep-water crustacean fishery operates, covers 4,000 km$^2$ off Portuguese mainland waters (including the EEZ) with depths from 200 to 700 m (Fig. 1). The geographical area is encompassed in ICES Subdivision IXa along the Iberian waters of the Northeast Atlantic Ocean. This region is located in a transition zone between temperate and tropical ecosystems (Briggs, 1974), characterized by high biodiversity associated with the different types of bottoms and relatively low abundance of commercially exploited marine species (Sousa *et al.*, 2005). The zoogeographic importance of this area has been recognized, representing the transition between north-eastern Atlantic warm-temperate and cold-temperate regions, which makes the Portuguese coast an area of great sensitivity to the detection of climate change (Teixeira *et al.*, 2016). Recent studies have shown that ecological responses to global warming are already visible along the Portuguese coast (Vinagre *et al.*, 2011; Teixeira *et al.*, 2014).

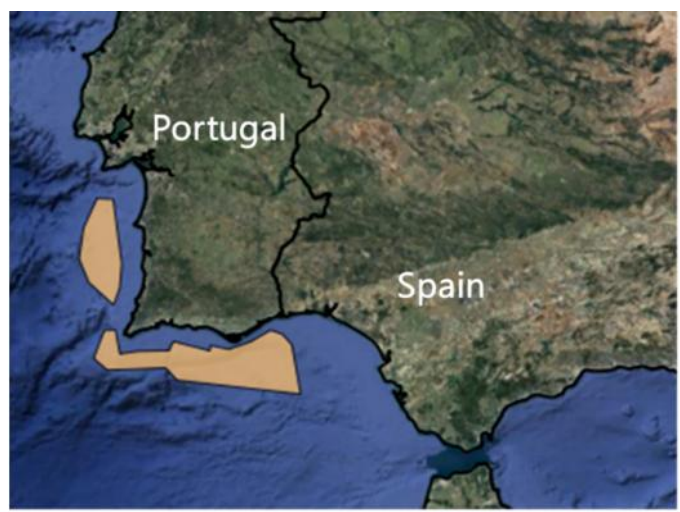

**Fig. 1.** Map of the study area (ICES Subdivision IXa) covering 4,000 km$^2$ with depths from 200 to 700 m where the crustacean trawl fishery operates.

The continental shelf is very narrow, varying between 7 and 28 km wide, followed by the relatively steep continental slope which is incised by several submarine canyons and trenches (Lopes and Cunha, 2010). The morphology reflects a geological evolution controlled by tectonic and sedimentary dynamics in this region (Hernández-Molina *et al.*, 2016). The only major output source, the Guadiana River, is located on the border between Portugal and Spain exhibiting strong annual and interannual fluctuations (Morais *et al.*, 2007; Cunha *et al.*, 2001; Sonderblohm *et al.*, 2014).

The physical oceanography is dominated mostly by the Gulf of Cadiz system, where the Mediterranean Outflow, a high salinity water mass flowing northward along the margin, plays an important role in controlling circulation dynamics (Relvas, 2002; Peliz *et al.*, 2009; Sánchez-Leal *et al.*, 2017). The area, under the influence of the Iberian system of the North Atlantic upwelling region, is subjected to seasonal hydrodynamic processes (Cunha *et al.*, 2001). The most productive waters are located close to the coast, while significant areas stretch over great depths with relatively levels of productivity (Borges *et al.*, 2001). In this region, the south-west winds are dominant especially during winter, both in frequency and in speed, while the north-west and south-east winds are moderate (Dias, 1988; Sonderblohm *et al.*, 2014).

Fisheries hold great tradition and socio-economic importance as a substantial source of employment along the entire coastal area (Borges *et al.*, 2001). The trawling fisheries comprise two fleets: one targeting fish such as Atlantic horse mackerel and hake, and cephalopods, whilst the second target crustacean species (Baeta *et al.*, 2009). Particularly important in terms of quantity of landings and/or value are sardine (*Sardina pilchardus*), horse mackerel, octopus (*Octopus vulgaris*), chub mackerel (*Scomber colias*), blue whiting, Atlantic mackerel (*Scomber scombrus*), hake, European anchovy (*Engraulis encrasicolus*), pouting (*Trisopterus luscus*), scabbard fish (*Aphanopus carbo*), seabreams (Sparidae), monkfish (*Lophius* spp.) and clams (Borges *et al.*, 2001). As in many other marine ecosystems, total landings trends have declined over the past decades with a number of species showing signs of overexploitation (Monteiro and Monteiro, 1999; Leitão, 2015).

## 3. Material and methods

### 3.1. The food web modelling approach

The food web model to represent the Southern and Southwestern Portuguese coasts (onwards SSWPT) was constructed using the *updated* Ecopath with Ecosim (EwE) software version 6.6 (Christensen *et al.*, 2008). In our study, we used the static module (i.e. Ecopath) to get a snapshot of the functional groups, according to similar ecological traits, and their interactions in a specific period, namely the year 2000. This year was selected because of the best data were available. The model is parameterized based on two master equations describing the production (Eq. 1) and consumption (Eq. 2) of each functional group:

Production = catches + predation mortality + net migration + biomass accumulation + other mortality (Eq. 1)

Consumption = production + respiration + unassimilated food (Eq. 2)

For each functional group, three of the four basic parameters [biomass (B), production (P/B) rates, consumption (Q/B) rates, and ecotrophic efficiency (EE)] are required and the fourth is left to be estimated by the model. Diet composition and catches are also needed. A detailed explanation of algorithms and equations are described in Christensen and Walters (2004) and Christensen *et al.* (2008).

### 3.2. Input data

In our study, 37 functional groups were considered to represent the SSWPT model including seabirds (FG1), marine mammals (FG2-3), turtles (FG4), fishes (FG5-19), cephalopods (FG20-21), invertebrates (FG22-32), primary consumers (FG33-34), primary producers (FG35) and detritus (FG36-37). The target and mainly discarded species by the fisheries were modelled separately (i.e. anglerfish FG7, hake FG9, mackerels FG10, horse mackerels FG11, megrim FG15, blue whiting FG18, rose shrimp FG22, Norway lobster FG23 and red shrimp FG24). The microbial loop was only indirectly parameterized as part of the detritus group due to lack of data, following other models (Torres *et al.*, 2013; Coll *et al.*, 2006; Tsagarakis *et al.*, 2010). A detailed description of each functional group is presented in the Supplementary Material (Table 1S).

Biomass estimates (t km$^{-2}$ year$^{-1}$) were estimated from bottom trawl surveys conducted in the study area using stratified and swept-area methods (PTGFS-WIBTS-Q4 targeting mainly fishes and PT-CTS (UWTV (FU 28-29) targeting crustaceans, ICES (2017b)). Both surveys are conducted seasonally on board the R/V Noruega since 1995 (in autumn) and 1997 (in summer) respectively (Azevedo *et al.*, 2014). Estimates of the abundance of vulnerable groups (i.e. seabirds and marine mammals) were taken from the 'MarPro Project' results. Quantitative estimates of phytoplankton, by means of chlorophyll-a values obtained from local sampling procedures were used to calculate the biomass (Table 1S). Detritus biomass was calculated empirically following the equation of Pauly *et al.* (1993) linking the production of detritus (D) with the average annual primary production (PP) and the average depth of the euphotic zone (E). Finally, biomass estimates for turtles, gelatinous plankton, benthic carnivorous feeders, macrobenthos and macrozooplankton were left to be estimated by the model using realistic EE values (Table 1S). Under mass-balance conditions, biomass accumulation and other export terms were assumed equal to zero.

According to Allen (1971), the production per unit of biomass (P/B, year$^{-1}$) was assumed as the total mortality (Z), which is the sum of fishing mortality (F, as the ratio between catches (Y) and biomass (B)), and natural mortality (M). Natural mortality for finfish species in absence of catch-at-age data was estimated from the empirical equation proposed by Pauly (1980):

Log M = – 0.0066 – 0.279 log L$_\infty$ + 0.6543 log K + 0.4634 log T (Eq. 3)

linking natural mortality (M), the curvature parameter of the von Bertalanffy growth function (K), the asymptotic length (L$_\infty$) calculated from Pauly (1984) where L$_\infty \approx$ L$_{max}$ / 0.95 and the mean water temperature (T in °C). Maximum length values (L$_{max}$) were obtained from the bottom trawl surveys. Growth parameters were taken from local studies or in absence from literature (Table 1S). The mean water temperature in the study area in 2000 was 13.5°C (Moreno and Santos, unpublished data). For most invertebrate species, secondary production ratios were taken from other models (Table 1S). Finally, P/Q values were assumed for seabirds, dolphins and benthic carnivorous feeders to let EwE estimate a P/B value (Christensen *et al.*, 2008).

Consumption per unit of biomass (Q/B, year$^{-1}$) for those groups with available information was calculated following the empirical equation of Pauly *et al.* (1990):

log (Q/B) = 6.37 – 1.5045 T' – 0.168 log W$_\infty$ + 0.1399 Pf + 0.2765 Hd (Eq. 4)

where W$_\infty$ is the asymptotic body weight (g), T' is an expression for the mean water temperature expressed using T' = 1000 / Kelvin (Kelvin = C + 273.15), and two variables expressing food type: Pf (1 for predators and zooplankton feeders and 0 for all others) and Hd (1 for herbivores and 0 for carnivores). W$_\infty$ was estimated by using length-weight relationships published in the study area or in absence from literature (Table 1S). Information related to the type of food was mostly taken from empirical data or from Fishbase (Froese and Pauly, 2018). For seabirds, consumption was estimated by using the empirical equation proposed by Nilsson and Nilsson (1976):

log (DR) = – 0.293 + 0.85 · log (W) (Eq. 5)

where DR is the daily ration (g) and W is the mean body weight (g) for each species. DR was converted to Q/B considering the biomass of each species and the time spent in the area. In the case of marine mammals, consumption was calculated from the empirical equation of Innes *et al.* (1987):

R = 0.1 · W$^{0.8}$ (Eq. 6)

where R is the daily ration (kg) and W is mean body weight (kg) for each species. R was converted to Q/B considering the biomass of each species (Table 1S). Due to the limited information available for crabs, benthic invertebrate carnivores and polychaetes, consumption was calculated as a function of body size following the equation of Finally, for those groups with no information provided in the area, the values were assumed from other models involving similar species (Table 1S).

The diet composition data were mostly compiled from local studies or from similar areas (Table 1S). Migratory species were taken into account by modeling a proportion of their diet compositions as import to the ecosystem following Torres *et al*. (2013).

In the SSWPT model we included catches (t km–2 year–1) from two separate fleets: crustacean trawl (OTB_crust) and fish trawl (OTB_fish). The official landings by species and fleet with reference to the year 2000 were provided by the DGRM (Portuguese Direcção-Geral de Recursos Marinhos, Segurança e Serviços Marinhos). Data of discards taken from local studies (Costa *et al*., 2008) were also included to give a more accurate total catch data. Unfortunately, estimates of illegal, unregulated, or unreported (IUU) landings were lack in the study zone.

### 3.3. Parameterization and mass-balancing

Prior balancing the SSWPT model, we checked that outputs of the model were within expected limits of a model ecologically and thermodynamically balanced (Christensen *et al*., 2008; Heymans *et al*., 2016): net food conversion efficiencies (P/Q [0.1–0.35], except for some fast growing groups), ecotrophic efficiencies (EE<1), respiration/assimilation (R/A [<1]), respiration/biomass (R/B [1-10]) for fishes and higher values for small organisms, and production/respiration (P/R [<1]) ratios (Table 2S SM). Unused consumption (U/Q) for those invertebrate groups mainly composed of filters, detritivores, and suspensivores feeders (i.e. FGs 29-34) was assumed to be 0.40 (Christensen et al., 2008). For the rest of the groups, the default value of 0.2 was maintained.

The mass-balancing was performed manually following a top-down strategy. The adjustments were performed according to their uncertainty degree, using the "Pedigree Routine". Initial results showed that the EE > 1 for most fish and invertebrate groups and we assumed that these groups may have had greater uncertainty in the biomass data. We therefore readjusted appropriate inputs parameters to achieve mass-balance following the same strategy applied in other ecosystems (Mackinson and Daskalov, 2008; Torres *et al*., 2013; Corrales *et al*., 2015).

We guide our balancing procedure following the PREBAL analysis (Link, 2010) to ensure that the model parameters followed some basic principles of ecosystem ecology. This analysis highlighted that some P/B and Q/B values had to be adjusted since they were too low or too high based on their trophic levels. After testing the model parameters with PREBAL, diets were slightly readjusted where needed, as in other Ecopath models (Coll et al., 2006; Torres et al., 2013). Cannibalism was also decreased in benthopelagic cephalopods, mesopelagic fishes, and crabs. Unfortunately, biomass estimates from the bottom surveys were not sufficient for the mass-balance. In the absence of catchability coefficients to correct biomass estimates, biomass for most groups of fish and invertebrates were left to be estimated by the model using a reasonable value of EE (EE values close but lower than 1) (Heymans *et al.*, 2016).

### 3.4. Pedigree index and quality of the model

The uncertainty associated to the input values in the model was quantified by using the pedigree routine (Christensen *et al.*, 2008). This index varies from 0 for low-quality models (i.e. input data estimated or taken from other models) to 1 for high-quality models (i.e. well-sampled and high-precision local input data). The pedigree index is averaged over all parameters and groups in the model to provide an index of the quality of the input data of the model.

### 3.5. Model analysis and ecological indicators

### 3.5.1. Ecosystem properties and trophic flows

A set of ecological indicators were used to assess the general ecosystem status and its stage of development and maturity *sensu* Odum (1971) and Christensen (1995). The overall ecosystem size in terms of flows is represented by the Total System Throughput (TST), calculated as the sum of the all the flows of consumption, exports, respiration, import and flows to detritus (Ulanowicz, 1986; Christensen and Pauly, 1993). In addition, the following indicators describing the system maturity are estimated: Total Primary Production/Total System Respiration (TPP/TR), Total Primary Production/Total Biomass (TPP/TB), and Total Biomass/Total System Throughputs (TB/TST) ratios. The Transfer Efficiency (TE) is defined as the fraction of the total flows at each trophic level (TL) that is either exported or transferred to other TLs through consumption. The mean TE is calculated as a geometric mean from the TE in trophic levels (TLs) II–IV (Christensen *et al.*, 2008).

Network analysis indices suggested by Ulanowicz (1986) were calculated. Ascendancy (A) provides information of the degree of development and maturity of an ecosystem as a measure of the average mutual information in a system, scaled by system throughput (Ulanowicz and Norden, 1990). The upper limit of A is called the 'development capacity' (C) and the difference between C and A is called 'system overhead' (O). The latter provides limits on how much the ascendancy can increase and reflect the system's 'strength in reserve' from which it can draw to meet unexpected perturbations (Ulanowicz, 1986). These indices are used as a measure of system stability, information, resilience and ecosystem maturity (Odum, 1969; Ulanowicz and Norden, 1990; Heymans *et al.*, 2007).

The cycling indices include the Finn's Cycling Index (FCI) representing the percentage of flows recycled in the food web (Finn, 1976), the Predatory Cycling Index (PCI) representing the percentage of recycling after the removal of detritus (Christensen *et al.*, 2008), and the Finn's Mean Path Length (MPL), which quantifies the mean number of nodes that energy inflow into the ecosystem passes through before exiting the network. In addition, the System Omnivory Index (SOI) and the Connectance Index (CI) correlated with ecosystem maturity and complexity of the food web (Christensen, 1995) were calculated. SOI is based on the average omnivory index (OI) which is calculated as the variance of the TL of a consumer's prey groups indicating predatory specialization (Pauly *et al.*, 1993). The CI is the ratio of the number of existing trophic links with respect to the number of possible links (Christensen *et al.*, 2008). The trophic structure was gathered into a Lindeman spine, an analysis of discrete TLs *sensu* Lindeman (1942) and proposed by Ulanowicz (1995). The system was aggregated into a linear food chain where import (on TL I only), consumption by predators, export, flow to the detritus, respiration, and throughput were calculated for each TL. The detritus box was separated from primary producers to show the amount of energy that is flowing through it. These flows were also represented by means of a flow diagram showing the trophic interactions between all groups within the ecosystem.

### 3.5.2. Trophic cascading and keystone groups

The Mixed Trophic Impact (MTI) analysis estimates the relative trophic impact (both direct and indirect) that a hypothetical increase in the biomass of a functional group would produce on the others within the ecosystem, including fishing activities (Ulanowicz and Puccia,1990). A positive or negative impact would mean an increase or decrease in the quantity of the impacted group. Further details or equations are well described in Christensen *et al.* (2008).

The index of Keystoneness (KS) evaluates the potential ecological roles of each functional group as keystones in the system. This index is a function of a group's trophic impact on other groups in the ecosystem and its biomass. Three methods proposed by Power *et al.* (1996), Librarato *et al.* (2006) and Valls *et al.* (2015) were compared. All methods use the relative overall effect calculated from the MTI against the KS and the contribution of each functional group to the total biomass of the food web.

### 3.5.3. Exploitation status of the fishery

Fishing impacts on the SSWPT ecosystem were assessed by analysing the mean trophic level of the catch (mTLc), the exploitation rates (F/Z), the relative consumption of total production representing the proportion of total production that is consumed within the system by all the functional groups, fishing mortalities (F), the gross efficiency of the fishery (GE, catch/net primary production), and the percentage of primary production required (PPR) to evaluate the sustainability of fisheries (Pauly and Christensen, 1995).

## 4. Results

### 4.1. The food web model analyzed by functional group

The 37 functional groups included in the SSWPT model were combined into four TLs ranged from 1.0 for primary producers and detritus groups to dolphins (TL=4.47) and anglerfish (TL=4.33), representing the top predators in the ecosystem (Table 1). The remaining fish groups TLs ranged from 3.08 (horse mackerels) to 4.21 (piscivores). Invertebrate groups were estimated to have a TL between 2.00 (benthic filter feeders) and 3.93 for benthic cephalopods. Seabirds which rely heavily on fishery discards probably had their trophic level underestimated (TL=3.04), as Ecopath assigns by default a TL = 1 to the fishery discards.

The highest flows to detritus corresponded to those groups ranked at the base of the food web. The highest FD values regarding fish groups were provided by small-sized and blue whiting in addition to the benthic groups. R/A ratio ranged from 0.33 to 0.99 t km–2 year–1, with the highest R/A values corresponding to top predators. OI showed that most groups were feeding on multiple TLs ranging from 0 to 1.445. The lower OI values were observed for benthopelagic fishes, horse mackerels and mesopelagic fishes, suggesting high specialization. These groups exert a high predation on zooplanktonic groups (macrozooplankton, micro- y mesozooplankton). By contrast, dolphins, hake, seabirds and minke whales presented the highest OI value with a large number of prey items in its diet. These results suggest more complexity on the upper part of the food web. These groups were followed by crabs, benthopelagic cephalopods, benthic sharks and Norway lobster, presenting a relatively wide trophic spectrum.

Regarding mortalities, most of the functional groups located in the upper part of the food web presented low predation mortalities (M2) as expected. In addition, the target species also showed low M2 values in particular red shrimp, even lower than the followers groups such as piscivores, and deep-sea fishes. Other natural mortality excluding predation (M0) was relatively low in most of the functional groups. However, phytoplankton, gelatinous plankton, and seabirds presented high M0 values.

**Table 1**: Basic estimates of the SSWPT model showing Trophic levels (TL), Biomass (t/km²/year), Production / Biomass (P/B (/year)), Consumption / Biomass (Q/B (/year)), Ecotrophic efficiencies (EE), and Production / Consumption (P/Q (/year)), Flow to detritus (t/km²/year), Net efficiency (NE), Omnivory Index (OI). Those values estimated by the model are shown in blue.

|    | Functional group name       | TL    | B     | P/B     | Q/B    | EE    | P/Q   | FD      | NE    | OI    |
|----|-----------------------------|-------|-------|---------|--------|-------|-------|---------|-------|-------|
| 1  | Seabirds                    | 3.039 | 0.017 | 3.083   | 61.659 | 0.002 | 0.050 | 0.260   | 0.063 | 1.191 |
| 2  | Dolphins                    | 4.473 | 0.030 | 0.070   | 7.000  | 0.210 | 0.010 | 0.044   | 0.013 | 1.445 |
| 3  | Minke whales                | 4.086 | 0.181 | 0.070   | 5.952  | 0.074 | 0.012 | 0.227   | 0.015 | 1.140 |
| 4  | Turtles                     | 3.649 | 0.000 | 0.170   | 2.480  | 0.400 | 0.069 | 0.000   | 0.086 | 0.205 |
| 5  | Benthic sharks              | 4.012 | 0.288 | 0.536   | 5.601  | 0.688 | 0.096 | 0.371   | 0.120 | 0.598 |
| 6  | Piscivores                  | 4.210 | 0.038 | 0.694   | 4.092  | 0.952 | 0.170 | 0.033   | 0.212 | 0.402 |
| 7  | Anglerfish                  | 4.328 | 0.082 | 0.554   | 3.586  | 0.602 | 0.154 | 0.077   | 0.193 | 0.119 |
| 8  | Rays and skates             | 3.970 | 0.196 | 0.354   | 4.095  | 0.428 | 0.086 | 0.200   | 0.108 | 0.240 |
| 9  | Hake                        | 4.112 | 0.180 | 1.190   | 4.162  | 0.902 | 0.286 | 0.171   | 0.357 | 1.318 |
| 10 | Mackerels                   | 3.282 | 0.711 | 1.140   | 6.118  | 0.950 | 0.186 | 0.910   | 0.233 | 0.355 |
| 11 | Horse mackerels             | 3.084 | 0.820 | 1.196   | 5.930  | 0.950 | 0.202 | 1.022   | 0.252 | 0.030 |
| 12 | Deep-sea fishes             | 3.466 | 0.766 | 0.503   | 6.286  | 0.950 | 0.080 | 0.982   | 0.100 | 0.204 |
| 13 | Benthopelagic fishes        | 3.123 | 1.239 | 1.080   | 5.413  | 0.950 | 0.200 | 1.408   | 0.249 | 0.021 |
| 14 | Mesopelagic fishes          | 3.138 | 2.666 | 2.590   | 9.070  | 0.950 | 0.286 | 5.182   | 0.357 | 0.055 |
| 15 | Megrim                      | 4.136 | 0.117 | 1.040   | 5.977  | 0.950 | 0.174 | 0.146   | 0.218 | 0.105 |
| 16 | Flatfishes                  | 3.472 | 0.100 | 1.364   | 8.730  | 0.950 | 0.156 | 0.181   | 0.195 | 0.120 |
| 17 | Large demersal fishes       | 3.434 | 1.377 | 0.703   | 5.654  | 0.950 | 0.124 | 1.605   | 0.155 | 0.285 |
| 18 | Blue whiting                | 3.477 | 1.437 | 0.746   | 5.431  | 0.926 | 0.137 | 1.641   | 0.172 | 0.250 |
| 19 | Small demersal fishes       | 3.295 | 1.480 | 1.203   | 10.100 | 0.950 | 0.119 | 3.080   | 0.149 | 0.114 |
| 20 | Benthic cephalopods         | 3.928 | 0.377 | 1.910   | 6.450  | 0.950 | 0.296 | 0.522   | 0.370 | 0.423 |
| 21 | Benthopelagic cephalopods   | 3.797 | 1.152 | 1.658   | 6.900  | 0.950 | 0.240 | 1.685   | 0.300 | 0.841 |
| 22 | Rose shrimp                 | 3.404 | 0.560 | 1.720   | 9.948  | 0.950 | 0.173 | 1.163   | 0.216 | 0.239 |
| 23 | Norway lobster              | 3.431 | 0.057 | 1.130   | 8.746  | 0.950 | 0.129 | 0.103   | 0.162 | 0.463 |
| 24 | Red shrimp                  | 3.540 | 0.055 | 1.350   | 8.063  | 0.950 | 0.167 | 0.092   | 0.209 | 0.180 |
| 25 | Shrimps                     | 2.998 | 2.847 | 3.210   | 9.220  | 0.950 | 0.348 | 5.706   | 0.435 | 0.416 |
| 26 | Crabs                       | 2.767 | 1.701 | 2.110   | 8.110  | 0.950 | 0.260 | 2.939   | 0.325 | 0.891 |
| 27 | Gelatinous plankton         | 2.895 | 0.096 | 13.870  | 50.480 | 0.400 | 0.275 | 1.762   | 0.343 | 0.179 |
| 28 | Benthic carnivorous feeders | 2.749 | 0.603 | 3.110   | 12.430 | 0.825 | 0.250 | 1.828   | 0.313 | 0.356 |
| 29 | Benthic filter feeders      | 2.000 | 2.615 | 0.800   | 6.500  | 0.950 | 0.123 | 6.904   | 0.205 | 0.000 |
| 30 | Bivalves and gasteropods    | 2.500 | 1.272 | 4.430   | 15.000 | 0.950 | 0.295 | 7.916   | 0.492 | 0.250 |
| 31 | Benthic worms               | 2.320 | 3.649 | 2.280   | 11.400 | 0.950 | 0.200 | 17.055  | 0.333 | 0.240 |
| 32 | Macrobenthos                | 2.100 | 2.881 | 15.620  | 52.120 | 0.950 | 0.300 | 62.324  | 0.499 | 0.090 |
| 33 | Macrozooplankton            | 2.100 | 1.804 | 20.410  | 50.940 | 0.950 | 0.401 | 38.589  | 0.668 | 0.090 |
| 34 | Meso- and microzooplankton  | 2.000 | 2.173 | 25.000  | 90.400 | 0.950 | 0.277 | 81.302  | 0.461 |       |
| 35 | Phytoplankton               | 1.000 | 5.043 | 220.491 |        | 0.203 |       | 886.363 |       |       |
| 36 | Detritus                    | 1.000 | 8.879 |         |        |       |       |         |       |       |
| 37 | Discards                    | 1.000 | 0.649 |         |        | 0.043 |       |         |       | 0.299 |

The total biomass supported by the ecosystem (excluding detritus and discard) was calculated as 48.14 t km$^{-2}$ in 2000. The largest part of the total ecosystem biomass was mainly represented by crustaceans (25.1%), fishes (23.9%), benthic invertebrates (16.9%) and cephalopods (3.2%). Primary producers occupying the lower trophic levels contributed to 10.5 % of the total biomass (5 t km–2). These results highlight the great importance of the demersal communities and benthic invertebrate producers in the area.

The quantification of the trophic flows among the functional groups are shown in Figure 2. It was observed that many groups from the pelagic compartment were consumed by groups in the benthic compartment. For example, the macrozooplankton (FG33) and benthopelagic and mesopelagic fishes (FG13 and 14) or macrobenthos (FG32) with mackerels (FG10) and horse mackerels (FG11). In addition, demersal groups such as piscivores (FG6), anglerfish (FG7) and hake (FG9) feed on groups from the pelagic compartment. The detritus was shown to be an important compartment in the SSWPT food web supplying most of the biomass and production in the demersal habitat (Fig. 2).

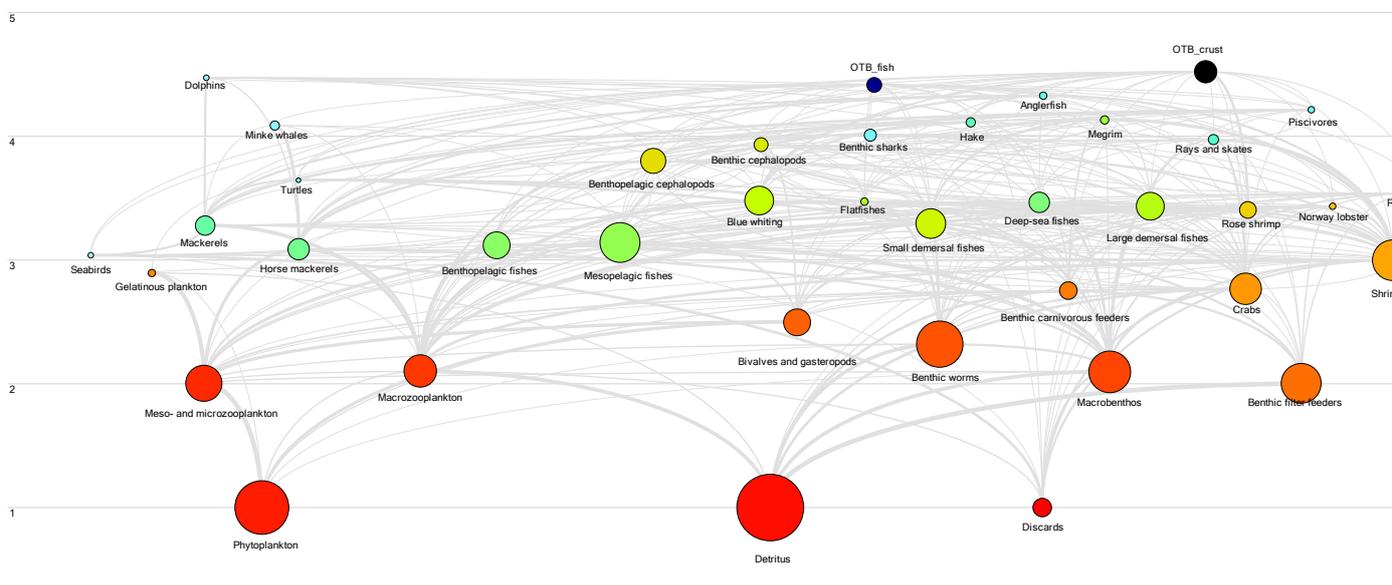

**Fig. 2.** Diagram of food web and trophic flows of the SSWPT ecosystem model. The size of each circle is proportional to the biomass of each of the functional groups.

### 4.2. Trophic flows

The Lindeman spine represented in Figure 3 revealed the importance of the TLs II and III in terms of biomass in the ecosystem as well as the later and TL IV in terms of exports and catches. Most flows were generated at the base of the food web. Primary production and detritus generated over 77% of the TST, underlying the importance of these groups moving the energy to the upper TL groups. TL II also contributed significantly to the major flows (TST = 16.8%).

The higher biomass concentration (excluding detritus) was found at TLs II and III composed of planktonic and benthic organisms, which together accounted for 79.6% of total biomass. Flows to detritus were mainly originated from TLs II. The most efficient trophic transfer (TE = 28%) occurred from TL II to TL III. The mean transfer efficiency (TTE = 22%) in the SSWPT system was higher than the general value of 10% estimated by Pauly and Christensen (1995). This means that on average 22% of the production of one TL became a production of its upper TL. The TE was mainly derived from primary producers (23%) and from detritus (21%). Finally, the exportation flows (i.e. catches) were mainly concentrated in TL III (0.786) followed by TL IV (0.425) and together with the high TE (22%) from TL III underline the strong exploitation by the fisheries on this TL (Fig. 3).

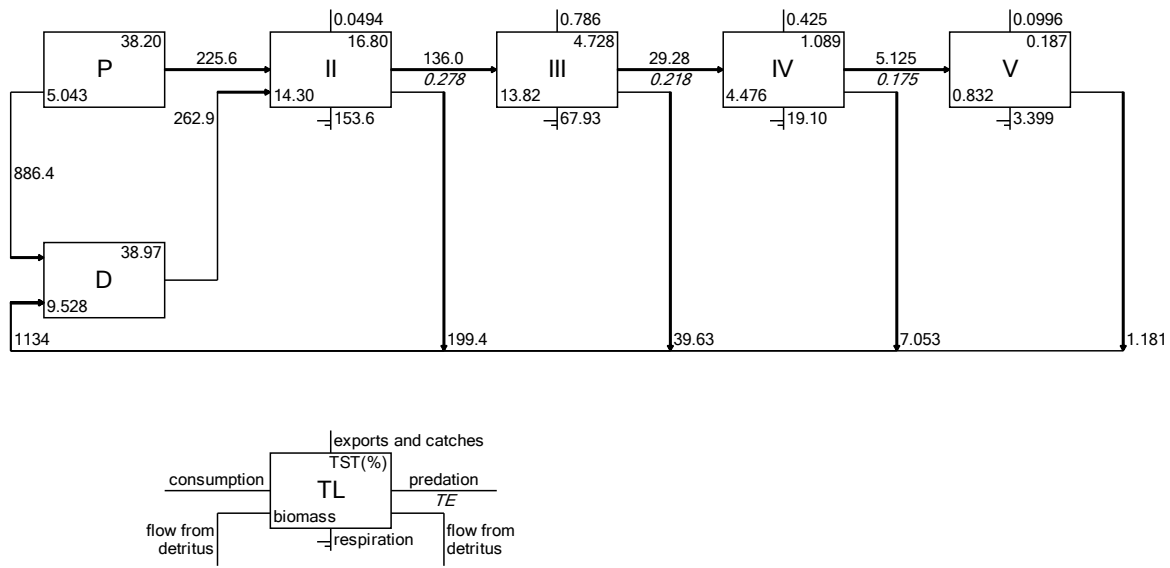

**Fig. 3.** Trophic flows of the SSWPT model organized by integer trophic levels (TL) in the form of Lindeman spine. TL I is split into primary producers (P) and detritus (D). Flows are represented in t·km$^{-2}$·year$^{-1}$.

### 4.3. Cascading effects and keystone species

The Matrix Trophic Index analysis (Fig. 4) ranged from +0.51, representing a positive effect of detritus on benthic filter feeders, to –0.95, revealing a strong negative effect of the fishery OTB_fish on dolphins. Overall, we observed that most of the functional groups would have a negative impact on themselves due to within-group competition for food resources and impacts on their main preys. Conversely, a hypothetical increase in biomass of the main preys would have a positive effect on their main predators. Numerous groups were impacted positively by the groups at the base of the food web such as detritus and macrozooplankton. Results from the MTI indicated that several functional groups showed a general low impact on the ecosystem, such as the target Norway lobster and red shrimp, as well as dolphins, turtles, piscivores, anglerfish, megrim and flatfishes. The MTI analyses also allowed us to identify a negative impact between the target rose shrimp, red shrimp and Norway lobster possibly due to competition for food resources.

Results from the keystoneness indexes (KS, Fig. 5) identified benthopelagic cephalopods, crabs, rose shrimp and macrozooplankton as keystone species/groups based on Libralato *et al.* (2006).

In the case of the index proposed by Power *et al*. (1996), seabirds, dolphins, minke whales, anglerfish, megrim and red shrimp were also highlighted as keystone. Following Valls *et al*. (2015), the groups identified were benthopelagic cephalopods, rose shrimp, minke whales and benthic carnivorous feeders. Accounting for the relative total impact, some groups with lower TLs were also identified as structuring groups (i.e. groups that have an important role in the ecosystem because have high biomass and high trophic impact) rather that unlikely keystone groups such as phytoplankton, shrimps and macrobenthos. These groups might indicate a possible bottom-up forcing.

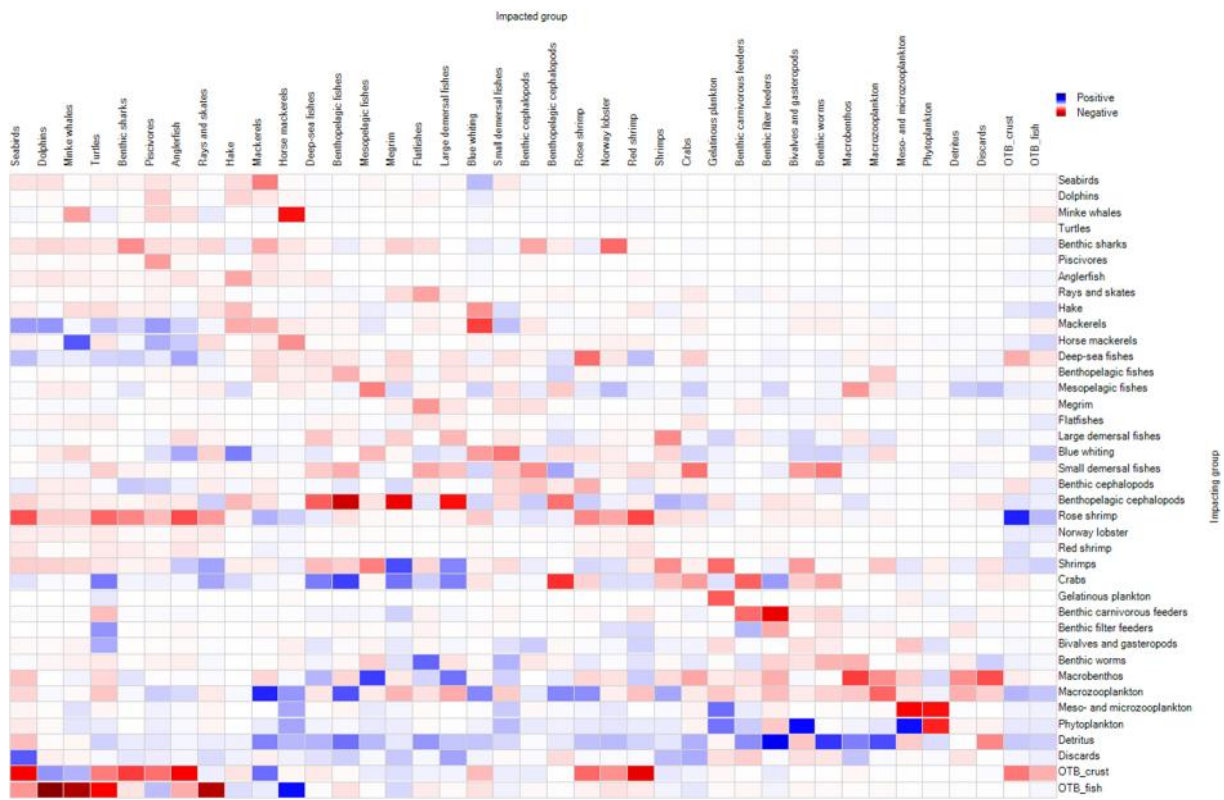

**Fig. 4.** Mixed Trophic Impact (MTI) analysis estimating the relative impact that a hypothetical increase in the biomass of a group would produce on the other groups within the ecosystem, including fishing activities. Negative (red) and positive (blue) impacts are represented for all functional groups.

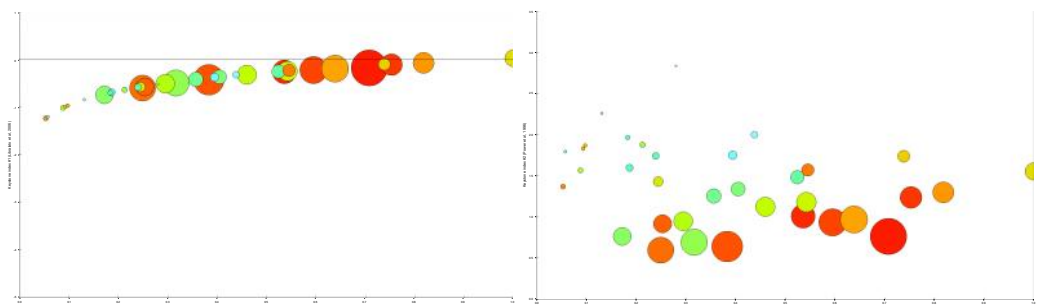

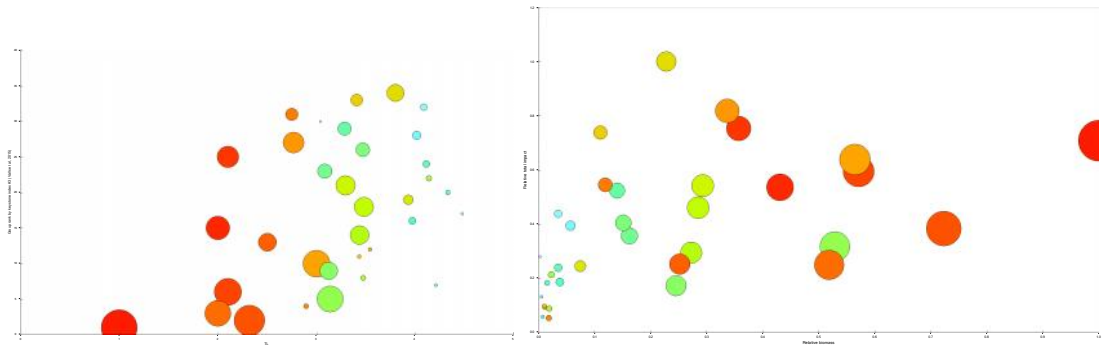

**Fig. 5.** Keystone Index (KI) analysis of the SSWPT food web. The size of the circles is proportional to the biomass of each functional group. KS indicators *sensu* (a) Librarato *et al.* (2006), (b) Power *et al.* (1996), (c) Valls *et al.* (2015) and d) overall impact.

### 4.4. Ecosystem status

Ecological indices calculated showed that the total system throughput was represented by consumption (33%), flows to detritus (32%), flows of all exports (23%) and respiration (12%) (Table 2). TPP/TR ratio (> 1), showed that the energy produced is more than four times higher than respired within the ecosystem (Christensen, 1995). The PP/B reflected a low level of biomass accumulation within the ecosystem compared to productivity. The information indices presented a high value of ascendency (89.7%). The system overhead in the SSWPT Portugal model was 10.34%. Regarding the food web complexity related indices, the omnivory index of the system was 0.22, the connectance index 0.31, the Finn's cycling index 4.82%, the Predatory Cycling Index (PCI = 0.35), and the Mean Path Length (MPL = 2.61).

*Table 2. Ecosystem indicators*

| Indicators | SSWPT | GoC | Cantabrian | Azores | Canarias | Wmedit | NWMed | Units |
|---|---|---|---|---|---|---|---|---|
| *Ecosystem Theory Indices* | | | | | | | | |
| Total system throughput (TST) | 2920.42 | 7734.85 | 7044.72 | 3587.91 | 7151.93 | 158.00 | 3758.03 | t·km−2·year−1 |
| Sum of all consumption (TQ) | 668.44 | 1946.87 | 2254.24 | 365.27 | 2684.88 | 51.64 | 897.27 | t·km−2·year−1 |
| Sum of all exports (TE) | 872.96 | 2233.67 | 1838.55 | 1470.90 | 1189.41 | 9.43 | 1088.08 | t·km−2·year−1 |
| Sum of all respiratory flows (TR) | 244.58 | 955.10 | 753.78 | 204.27 | 1009.39 | 20.17 | 279.55 | t·km−2·year−1 |
| Sum of all flows into detritus (TFD) | 1134.44 | 2599.22 | 2198.16 | 1554.47 | 2268.25 | 65.98 | 1493.14 | t·km−2·year−1 |
| Sum of all production (TP) | 1298.94 | 3704.44 | 3488.12 | 1763.11 | 3052.56 | 50.78 | 1599.93 | t·km−2·year−1 |
| Calculated total net primary production (NPP) | 1112.00 | 3187.67 | 2838.24 | 1675.16 | 2192.65 | | 1366.1 | t·km−2·year−1 |
| Total primary production/total respiration (TPP/TR) | 4.55 | 3.34 | 3.77 | 8.20 | 2.17 | | 4.89 | |
| Net system production (NP=TPP-TR) | 867.43 | 2232.57 | 2084.46 | 1470.90 | 1183.26 | | 1086.55 | t·km−2·year−1 |
| Total primary production/total biomass (TPP/TB) | 28.80 | 39.84 | 42.07 | 67.73 | 8.65 | | 32.00 | |
| Total biomass/total throughput (TB/TST) | 0.01 | 0.01 | | 0.01 | 0.04 | 0.03 | | year−1 |
| Total biomass (excluding detritus) (TB) | 38.61 | 80.02 | 67.47 | 24.70 | 253.57 | 3.92 | 42.69 | t·km−2 |
| Mean transfer efficiency (TTE) | 22.00 | 14.90 | | | 15.80 | 15.80 | 14.30 | % |
| Connectance Index (CI) | 0.26 | 0.25 | | | 0.15 | | | |
| System Omnivory Index (SOI) | 0.35 | 0.18 | | 0.22 | 0.34 | 0.29 | 0.19 | |
| Ecopath pedigree index | 0.55 | 0.63 | | 0.53 | 0.64 | 0.54 | 0.62 | |
| *Fishery Indices* | | | | | | | | |
| Total catch (TC) | 1.38 | 4.55 | 6.42 | | 4.55 | | 4.18 | t·km−2·year−1 |

| | | | | | | | |
|---|---|---|---|---|---|---|---|
| Mean trophic level of the catch (mTLc) | 3.49 | 3.32 | | 3.95 | 3.43 | | 3.13 | |
| Gross efficiency (catch/net p.p.) (GE) | 0.001 | 0.00 | | | 0.00 | | 0.00 | |
| Primary production required to sustain the fishery (PPR, considering PP) | 1.02 | 12.97 | | | 21.22 | | 12.08 | % |
| Primary production required to sustain the fishery (PPR, considering PP + detritus) | 12.61 | 16.45 | | | 46.92 | | 17.36 | % |
| *Cycling Indices* | | | | | | | | |
| Predatory cycling index (PCI, of throughput without detritus) | 0.46 | 8 | | | 7.89 | | | % |
| Throughput cycled (excluding detritus) | 3.01 | 2.14 | | | 206.46 | | | t·km−2·year−1 |
| Finn´s cycling index (FCI, of total throughput) | 0.96 | 3 | | | 12.6 | 4.2 | 9.12 | % |
| Finn´s mean path length (MPL) | 2.61 | 2.43 | | | 3.253 | | 2.75 | |
| *Information Indices* | | | | | | | | |
| Ascendency (A) | 89.66 | 41.1 | | | 25.5 | | | % |
| Overhead (O) | 10.34 | 49.2 | | | 74.5 | | | % |
| Capacity ( C) | 10875 | 25810 | | | 31637.9 | | | Flowbits |

## 4.5. Fishing impacts

In the SSWPT model, the total catch in 2000 was 1.38 T/km$^2$ (Table 2). The mTLc was 3.49, matching the TLs of the dominated groups in landings, which corresponded to rose shrimp, followed by red shrimp, hake, horse mackerels and Norway lobster. By contract, the most discarded group was blue whiting, followed by mesopelagic fishes, benthopelagic fishes, mackerels, benthic sharks and large demersal fishes.

Exploited functional groups showed high values of exploitation rates (F/Z) such as red shrimp (0.77), piscivores (0.60), anglerfish (0.60), Norway lobster (0.54), benthic sharks (0.52), rose shrimp (0.42) and rays and skates (0.40). Fishing mortality values (F) were also high (> 0.5) for the three exploited crustacean species, including red shrimp (1.04), rose shrimp (0.72) and Norway lobster (0.61), indicating that more than 50% of total mortality in the group was due to fishing (Table 1).

The primary production required to sustain the fisheries (PPR%) when considering the detritus and the primary producers was estimated at 12.61 % in the SSWPT ecosystem in 2000. The major PPR fractions were to sustain mainly the catches of blue whiting, rose shrimp, benthic cephalopods, and mesopelagic fishes. The gross efficiency of the fishery showed a value of 0.001.

The MTI analysis showed the impact of the two trawling fisheries on the ecosystem. Overall, the two fleets presented negative impacts on several groups with the largest negative impacts found on its main target species. In particular, the crustacean trawl fleet had strong negative impacts on the target red shrimp rose shrimp, and Norway lobster, but also on top predators such as benthic sharks, anglerfish, piscivores and rays and skates related to the high rates of discards of these groups by this fishery. Also, it was identified a strong negative impact on seabirds showing indirect impacts on this group. In general, the fish trawl fleet showed the highest negative impact on hake via direct mortality and indirectly on marine mammals possibly due to competition for same resources.

## Acknowledgments

This research has received funding from the European Commission's Horizon 2020 Research and Innovation Programme under Grant Agreement No. 634495 for the project Science, Technology, and Society Initiative to minimize Unwanted Catches in European Fisheries (MINOUW).